\documentclass[conference]{IEEEtran}
\IEEEoverridecommandlockouts
\usepackage{cite}
\usepackage{amsmath,amssymb,amsfonts}
\usepackage{algorithmic}
\usepackage{graphicx}
\usepackage{textcomp}
\usepackage{xcolor}
\def\BibTeX{{\rm B\kern-.05em{\sc i\kern-.025em b}\kern-.08em
		T\kern-.1667em\lower.7ex\hbox{E}\kern-.125emX}}
\begin{document}
\title{Leveraging Data Scientists and Business Expectations During the COVID-19 Pandemic}

\author{\IEEEauthorblockN{Wellington Rodrigo Monteiro\textsuperscript{\textsection}, M{\'a}rcio Leandro do Prado\textsuperscript{\textsection}, Gilberto Reynoso-Meza}
	\IEEEauthorblockA{\textit{Pontif{\'i}cia Universidade Cat{\'o}lica do Paran{\'a}}\\
		Curitiba, Brazil \\
		wellington.monteiro; marcio.prado; g.reynosomeza@pucpr.edu.br}
}

\maketitle
\begingroup\renewcommand\thefootnote{\textsection}
\footnotetext{Both authors were the technical and team leaders of the team presented in this manuscript at the time it was written and contributed equally to this research.}
\endgroup


\begin{abstract}
  The COVID-19 pandemic presented itself as a challenge for separate societal sectors. On the information technology (IT) standpoint, it does include the maintenance of the infrastructure required to hold collaborative activities that went to happen online; the implementation of projects in a scenario of uncertainty; and keep the software engineering and information security best practices in place. This article presents the context of a data science team organized as a skunk works group composed of professionals with experience in both the industry and academia, located in an IT department working with a team of seasoned data engineers. At the time the pandemic started, the relatively new data science team was positioning itself as a Center of Excellence in Advanced Analytics. With the pandemic, it had to keep up with the expectations from the stakeholders; manage current and upcoming data science projects within the methodology practiced in IT; and maintain a high level in the quality of service delivered. This article discusses how did the COVID-19 pandemic affected the team productivity and its practices as well as the lessons learned with it.
\end{abstract}


\maketitle

\section{Introduction}

Company B (name redacted due to the company policies) is one of the largest food producers in the world with more than 90000 employees and operating in more than 130 countries. The Information Technology (IT) department is comprised of several teams according to their area of expertise such as, but not limited to IT architecture; infrastructure and network; front-end and back-end developers; information security; business support; mobile and emergent technologies; analytics and Artificial Intelligence (AI). The business support teams in this department are also structured according to their knowledge domain (e.g., finance, engineering, production, agriculture) and their support levels (project implementation and daily support).

In the late 90s the company was one of the pioneers in South America in implementing the ERP solution it still currently uses to control its operations. Due to the particularities and complexities of the Business activities (e.g. large and broad supply chain from animal feed to food sales; diverse health certificates depending on the states, countries and trade blocs; handling of livestock; monitoring the consumption of utilities in separate plants and at separate levels; plants and production lines with varying complexities; volatility of the global commodity markets; lab analysis follow-ups; custom marketing and sales strategies; extensive road network in use to sell goods in the internal market), the ERP has extensive customization levels in order to meet all the Business operations, Audit and IT sustainability requirements.

The transactional data stored in the ERP enabled the usage of accurate, integrated reports and quicker decisions in the Business Warehouse (BW) solution from the same vendor. In order to support this solution, a new team was first assembled in 1998 known at the time as the Business Warehouse (BW) team. It had three main activities: to ensure all month-end closure activities were executed flawlessly and respecting all the IT Governance best practices and Audit rules; to occasionally reprocess transactional data into the BW systems if any data error in the past had surfaced; and to provide simple mathematical simulations based on the BW data to the Business teams. All this criteria followed the trade-off between high query performance and low view maintenance costs \cite{Theodoratos1997}.

From this time and up to the early 2010s the company used a ERP-centric, ERP-first approach where most of the transactional data was held in the ERP except for some special use-cases (e.g. production plan optimization, customer relationship management (CRM), sales order creation on-the-go from the salespersons driving to customers such as bakeries, groceries, supermarkets and cafes). Software engineering processes including software lifecycle management, feasibility and requirement studies, software maintenance, IT governance, internal and external audit, change and incident management matured as well as the data quality and reliability. With it, the requirement for reporting dashboards grew across different business units. In 2012, the team was renamed as the \textit{Business Intelligence} (BI) team. Organically growing on the aforementioned activities, it also provided strategic reports and static dashboards to several Business teams, ranging from Procurement to Sales; from Production to Logistics; from the Farms to the Factories and Distribution Centers.

Up to 2016, the number of dashboards grew to the hundreds. In this year, the team had been renamed to \textit{Business Analytics \& Performance} and commonly referred as \textit{Analytics}. The activities were segregated between providing strategic dashboards -- intended to be used by the high management for strategic decisions -- as well as tactical dashboards intended to daily operations under the guidance of the lower management. At this time, the team was deeply focused on activities related with descriptive analytics and split between data engineering and data analysis roles. In parallel, the whole department was moving towards a path where the ERP was not to be considered anymore as the ``one-stop-shop'' to most of the IT-related needs, but instead also integrating with other specialized systems depending on the Business needs (e.g., laboratory information management systems (LIMS); CRMs; livestock management systems; and HR specialized systems) following Enterprise Architecture (EA) principles and a cloud-first, data-driven approach.

With it, the data requirements went from consuming tabular, transactional data to other data types from sources such as Internet of Things (IoT) sensors, image and video feeds and telemetry data as both unstructured and semi-structured data from internal and external sources. Furthermore, the overall amount of data being created also grew consistently. Consequently, the number of requests for deeper data analysis and predictive analytics multiplied. On the other hand, there was also an increasing number of upcoming requests for changes in already-deployed dashboards and diverging discussions from separate Business units on what was the correct number for any given metric or key performance indicator (KPI) since each area could have different interpretations on the same data based on their own experiences. Therefore, in the second half of 2018 the IT architecture intended for Analytics moved towards a ``self-BI'' orientation, where all the upcoming tactical dashboards were intended to be created and maintained by the Business teams with the Analytics team (in the IT department) in charge of ensuring the underlying data is reliable and readily accessible as well as guaranteeing the users could only see the data they were intended to see and with data filters respecting the shared usage. Meanwhile, the overall IT project management direction moved towards the mainstream adoption of the Agile methodology.

At this time, the concept of \textit{Advanced Analytics} had been introduced to the team. Organized since its inception as an internal ``skunk works'' team \cite{biron2020}, it was founded under four desiderata: first, to provide data science solutions taking advantage of already being part of the team in charge of overseeing and curating data; second, to spearhead advanced analytics initiatives within the company looking to balance state-of-the-art techniques and EA principles already in place for a sustainable environment under an IT standpoint; third, to form a highly-skilled team from professionals with experience in academic initiatives but also with Business and/or IT industry expertise; fourth, to conduct experiences and introduce new techniques should they bring competitive advantage to the company. Following this logic, the year of 2019 focused on finding potential use-cases, experimenting with them, and probing industry solutions intended for data science purposes readily available in the contracts already in place with the IT vendors.

As part of the Analytics team, the data science group was not assembled separately from the already existing team nor with a separate chain of command -- in fact, the idea was to leverage the expertise of the data engineers and workflows already in place in the team, as well as having an easier access to data and the people holding the knowledge on them. Besides, the idea was also to foster the usage of advanced analytics techniques and tools among the team, slowly distancing from legacy activities inherited from the past years with less perceived value such as visual changes in dashboards and manual data load for situations where automation is not possible -- activities that could be delegated to the Business teams.

The previously mentioned data science use-cases were well-received by the Business teams and, with it, the demand for data science projects grew for 2020. In this year, the Center of Excellence (CoE) in Advanced Analytics had been formed following the same desiderata mentioned earlier as well as positioning itself as the source of data science best practices and new implementations in the company. Due to the size of the company and its extension, it proved as a challenge. This challenge was further expanded by the COVID-19 pandemic in a context where the company value chain is concerned with keeping the continuous delivery of perishable goods in a reliable fashion to the consumers keeping the same levels of quality and safety. Therefore, the workload for all the teams -- both in and out of IT -- multiplied. While most teams were located across several geographic locations, many activities were face-to-face such as Agile squad meetings, workshops, and team meetings. At the same time, the company was focusing on evangelizing Business teams to work with Agile methodologies and virtual interactions.

Considering this scenario, this manuscript proposes a discussion on how the team -- more specifically, the CoE in Advanced Analytics -- managed its ongoing activities and the expectations of the Business areas during the COVID-19 pandemic. It is divided in four sections: the first and current section introduces the reader to the history and context of the team. The second section covers the team situation shortly before the start of the pandemic. The third section shows the challenges experienced by the team during the pandemic. The fourth and last section has the concluding remarks of the authors of this manuscript.

\section{Situation before the pandemic}
\subsection{Data Engineers}
In the start of 2020, the team had two roles divided across about twenty-five full-time employees (not including contractors): data engineers and data scientists. The work happened in a cloud infrastructure including databases, data lakes, data warehouse systems, business intelligence software, serverless applications for web scraping tools, cloud AI systems, private Git repositories and integrations with other company systems such as the ERP. Several IT governance processes such as change management and incident management were mature and fully operational for the data engineers -- mostly because many of the analysts were already seasoned professionals and the team had evolved organically since its inception. Furthermore, these professionals had several contacts during the years with different global vendors and Audit teams.

Most of the team filled the role of data engineers and were already in the team before the introduction of advanced analytics in 2018. They are focused on activities such as creating and maintaining data models, architecture, and pipelines as well as their ingestion, transformation and integration reflecting the notation found in \cite{saltz2017}.

They either were experienced Business professionals with heavy knowledge of their previous activities, understanding Business processes, practices, specific terms, and daily activities, but moving to IT for also displaying technological proficiency; or were professionals with an academic background in IT with an acquired expertise of the Business practices and requirements. The data engineers are organized into smaller ``fronts'' to provide specific, tailored services to areas such as Sales, Logistics, Agriculture and Procurement. Each one of these fronts has from two up to four professionals working in parallel. This organization was organically created within the team during the years and proved to be hugely successful due to some reasons, as follows: first, the Business teams saw in these fronts as reliable, focal points which understood their lexicon, processes, and urgencies. Second, the fronts could rapidly share knowledge and re-prioritize their activities with more freedom. Third, the fronts were organized in the office in a way the whole team was close, and the engineers for the same front were even closer. Fourth, the engineers could better organize their schedules to cover medical leaves and vacation periods while keeping the same quality of service.

As previously mentioned, in the upcoming years the demand for Agile initiatives increased within the department and, later, to the whole company. In this scenario, the organization in knowledgeable fronts enabled the possibility of independently assigning team professionals for parallel Agile squads while ensuring the other tasks were still being worked on by other engineers for the same front.

\subsection{Data Scientists}
The data scientists, on the other hand, started moving to the team in the second half of 2018. Up to the second half of 2019, it was composed of a single data scientist, but grew to be around one third of the Analytics team by the start of 2020. Following the description of \cite{saltz2017}, they are more versed in machine learning, dataset handling, data science techniques and data analysis which, in turn, includes the adoption of statistics.

These scientists were either seasoned IT professionals working for other teams in the same department and who were independently pursuing Master's and Doctorate degrees in AI, or professionals who had already attained these degrees applying advanced analytics techniques in the industry, but coming for other companies where they could not apply these skills in their previous positions. Therefore, the data scientists were mainly professionals who had merged scientific and academic knowledge with the professional experience instead of severely lacking experience in the former or latter.

The data scientists also were split between two separate profiles: the \textit{``data scientists (machine learning)''} -- data scientists holding bachelor's degrees in IT and holding a solid knowledge on IT practices and theory; and the \textit{``data scientists (statistics)''} -- data scientists from Mathematics, Statistics and other domains of knowledge where these areas are applied. These scientists can handle statistical analysis and understand how to handle separate problems using quantitative techniques.

These profiles meet the roles considered as part of an effective data science team mentioned in \cite{bavskarada2017unicorn}. The domain experts and data engineer roles are filled by the data engineers in this team; and the statisticians and computer scientists are filled by the two data scientist profiles. As it will be discussed in the following sections, the communicator role is interchangeably filled by both data engineers and scientists in the team.

From an internal organization standpoint, the data scientists adopted a more flexible approach by organizing themselves as a ``skunk works'' team covering all the fronts in charge by the data engineers in the team. A skunk work is a team with a high degree of autonomy, capable of generating innovation by being less constrained by processes in place for other teams and an alternative for organizations that are expected to be primarily focused on delivery and execution with quality instead of innovation (except for this kind of team) \cite{biron2020}. The high degree of autonomy of the team was partly justified by the highly skilled profile of these scientists and partly by a requirement of experimenting several processes and techniques that could be more effective to the organization. Furthermore, since these processes and techniques were often out of the expertise handled by other teams, there was also a possibility of experimenting with what, until then, was mostly unknown by the organization.

Instead of using the waterfall or Agile approaches, the data scientists went to adopt the Cross-Industry Standard Process for Data Mining (CRISP-DM) methodology. CRISP-DM is one of the most used methodologies in data science projects and is based on an approach focused on following a list of stages from data to knowledge \cite{martinez2019crisp}. Its process is represented in the Fig. \ref{fig:crispdm}. Foreseeing a need to better explain how the methodology works to the stakeholders, the team attempted to get closer to Scrum which, in turn, was being introduced to several Business teams to be used in non-IT projects. While Scrum is composed of roles such as the Product Owner (PO), Scrum Master and the developers \cite{schwaber2011}, CRISP-DM does not cover roles \cite{chapman1999}. However, the team included within its data science projects the following profiles:
\begin{itemize}
	\item PO (required): lending the name from Scrum, this role is occupied by a Business key user -- often a senior analyst or a specialist -- possessing the knowledge required to understand the data relationships in their specific department and their context. They are able to communicate to their leadership who had never worked in a data science project before what are the current data science project activities and the expectations for the upcoming phases and cycles;
	\item Business support (optional): other users who might help in extracting, understanding, and explaining data and their meaning. They are often holding more junior positions;
	\item Data scientists (required): data scientists of the Analytics team in charge of developing the data analysis and, when applicable, creating predictive models;
	\item Data engineers (required): data engineers of the Analytics team in charge of aiding the data discovery, understanding, and processing steps.
\end{itemize}

\subsection{Building Trust}
As noted in \cite{bavskarada2017unicorn}, the adoption of data science is only useful if it can affect organizational change. Therefore, effectively communicating useful results, understanding the Business problems, and positively getting the attention of the stakeholders rapidly is paramount.

In the Analytics team, this role is shared between both data engineers and data scientists. The data engineers are valued by the Business teams for their valuable knowledge, reliability and efficiency built over the years. From this position, they can share discourses with the Business teams and have a common ground. The stakeholders already can set expectations on targets such as deliverables and deadlines based on previous experiences.

The data science projects, however, presented themselves as a challenge. The stakeholders were unaware of the CRISP-DM methodology and did not possess the same prior experience as the data engineering projects. In parallel, due to the size of the company, startups and large vendors often offered AI services to several Business teams. However, in the sales discourse, these companies often illustrated their solutions as easy, simple to create and with huge performance gains attempting to include terms often mentioned in the broad public such as ``deep learning'', ``neural networks'', ``predictive analytics'' and ``AI'' demonstrating with simpler, toy datasets such as \textit{cars} \cite{krause2013}. This speech -- part of a hype built by technology companies investing in creating this market \cite{gandomi2015} --  often hid issues such as: adapting and complying with EA, IT Governance and Information Security; explaining that the algorithm performance is hugely tied to data and most of time is spent by understanding and sanitizing the datasets; understanding the usage of complex AI architectures such as deep neural networks (DNNs) are often not the best alternatives for real business cases; some of the solutions built might not be cost optimized or even be oversized and overpriced at the expense of the lack of a buyers' knowledge in data science.

\begin{figure}[h]
	\centering
	\includegraphics[width=\linewidth]{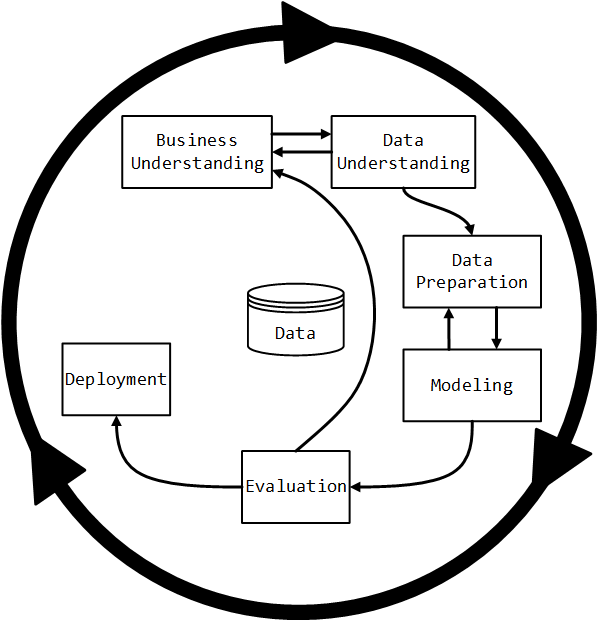}
	\caption{CRISP-DM lifecycle}
	\label{fig:crispdm}
\end{figure}

Adding to the external pressure from the vendors, there is the internal pressure maintained by the company departments. The IT department, as the natural bulwark of software innovations, is expected by the other departments to support, understand, and proactively propose technical solutions which, in turn, includes data science. The other departments, also including professionals capable of technically understanding some of the intersecting subject matters which compose data science such as mathematicians, statisticians and software developers allocated in these departments, could develop code conforming to their own needs and without worrying about automation, maintenance or validating software requirements. However, the code created by the Business teams is often executed locally and commonly without integration with other software and databases. Therefore, the data scientists in the Analytics team had to understand the varying degrees of software development knowledge and expectations of the Business teams to build trust -- both by trusting that the data scientists in IT had all the knowledge required to help them in the best way, and by trusting that the methodology and key points brought by them are part of the untold story offered by the vendors.

To achieve this end, the data scientists followed some criteria, such as:
\begin{itemize}
	\item When first discussing a possible data science project with the stakeholders, the conversation focused on interdependence, people, and results, all of them being key values of the company.
	\subitem \textit{Interdependence} is understood as important for building a reliable data analysis and algorithms based on the knowledge of Business experts in the company and data shared. Often, the data scientists mention that an algorithm thoroughly tested by the stakeholders build trust: if seldom an algorithm in real-world use cases have 100\% accuracy, as soon as a mislabeled case appears, if the data science solution is presented as a black-box solution and if the stakeholders do not consider themselves as people that build together the solution, they will in turn reject the notion the solution meets their demands. However, if they are involved since the first steps of the project and help validating the data and understand how does the solution work, they will also feel themselves (with all the due justice) as co-owners of the solution instead of just being users.
	\subitem \textit{People} is understood as important for building a partnership based on knowledge sharing -- the data scientists understand it is important to not only understand together how does the data structures are built and identify Business cases or data anomalies, but also to keep the knowledge within the company and shared between the teams. The knowledge shared and kept within the company is considered by it as one of its key values. Algorithm solutions such as Explainable AI (XAI) techniques are also helpful in showing how does machine learning algorithms work and foster the construction of trust and transparency for the users \cite{samek2019towards,weller2019transparency}.
	\subitem \textit{Results} cover the notion of having quantitative metrics showing both the possible gains by implementing a data science project and measuring the gains during and after the project. It also does cover sustainable solutions (both considering the environment and IT operations), efficiency, excellence, and innovation. Considering the CRISP-DM process shown in the Fig. \ref{fig:crispdm}, it is useful to set objectives of the project (Business Understanding step) closer to the reality and in a way to quickly retrieve data and measure results.
	\subitem Considering the stakeholders, \textit{results} may also be considered as understanding the project deliverables. By displaying and explaining how does the project work based on the Fig. \ref{fig:crispdm}, the data scientists are able to show that the process is transparent and composed of non-linear paths. While a go-live date is often difficult to inform, data scientists show the project objectives may change during the project depending on the data and the steps depend on the data and people. One of the examples mentioned a case where the steps of Business Understanding and Data Understanding went back and forth for some weeks as the data granularity and availability was under study. Therefore, a real-time predictive algorithm could be turned into an algorithm which processed data twice per day which, in turn, could pave the way for a different set of expectations and opportunities. On another example, the data preparation step took up to six months and the modeling step took around one day while, on a third example, the inverse had happened. The data scientists also mentioned that for all the covered examples the Business teams could better understand how their data worked and could prepare for future projects to increase data quality. On a specific case, during the data preparation step the Business teams discovered a possible hotspot on their data and could preemptively prepare an action plan to avoid any issues in the future. These cases helped to show the Business teams that the lifecycle follows separate iterations, and its progress is measurable for all the cycles with potential deliverables for all the covered phases, comparable to the concept of iterative and transparent development from Scrum \cite{ramin2020}. This comparison also helped foster trust and control anxiety.
\end{itemize}

\subsection{Advanced Analytics}
Understanding the need to avoid distancing from the \textit{analytics} term, but also needing to present itself as a team capable of supporting operations that could be considered by the stakeholders as ``AI'', ``Data Science'' and ``Statistics'', the data science skunk works group named itself as the \textit{Center of Excellence in Advanced Analytics} (CoE AA) with the department stewardship. The term \textit{advanced analytics} was chosen since it referred to both statistical and machine learning models as well as optimization, forecasting and similar problems \cite{franks2012}. The word \textit{advanced} also implied the use of new techniques the stakeholders could be unaware of.

Many of the activities of the data engineers are also considered as part of advanced analytics. However, its usage within the company refers to the application of techniques intended to be used by data science projects only. While the near-term idea is to move the whole team under the CoE name, currently it does include only the data scientists in the skunk works group and the data engineering technical lead.

\section{Situation during the pandemic}
The pandemic brought a series of additional challenges to the team. When all administrative activities moved to the work-from-home format the CoE AA was relying on face-to-face contacts by explaining concepts in whiteboards located in huddle rooms; drawing concepts in paper notebooks with the stakeholders; and meeting new contacts in the corridors within the office. In the office where the Analytics team is located, some Business teams such as Agriculture, Logistics and Procurement as well as the other IT teams in the same department. Key users from other offices in teams such as Sales, Finance and from the factories often traveled to the same office for some projects. It is essential to highlight the fact that both the headquarters and this office in specific were in South America, where the work culture heavily emphasizes physical contact and face-to-face meetings. The data engineers often discussed issues and new alternatives in their desks in an open-office environment. This situation favored the introduction of the data scientists and paved the way for new opportunities through networking.

The IT department was implementing a myriad of projects to modernize operations with a heavy emphasis on Industry 4.0 features such as automation, IoT, real-time data and AI \cite{schwab2017}. With it, being in the same physical location as the other IT teams was strategic to quickly reach project managers, understand current initiatives and negotiate the participation in new projects. While it is relevant to highlight the IT higher management supported the CoE, due to the soaring amount of parallel initiatives the other analysts often did not notice or postponed a contact with the CoE.

With the pandemic, the team had to ensure all the current projects were running and keeping its purpose of expanding its operations, understanding the needs of current and new stakeholders, train new employees in the team and keep fostering innovation.

\subsection{Productivity}
The productivity in the first weeks suffered a hit due to the uncertainty in short term. The local government in the large cities (where the offices are located) placed heavy restrictions in the first days affecting public transportation, commerce, and offices. Since the company works with food products, the uncertainty of the sales prospects in the upcoming months affected all Business teams since the pandemic could affect not only the balance between exports and imports, but the currency exchange rates and the balance between sales for food service and end consumers. The uncertainty required an unprecedented workload for the data engineers for new data handling and visualization requests. On the other hand, discussions for new data science projects remained on hold due to the other urgent issues.

After the first weeks, the urgent issues were addressed by the teams and the data science projects resumed. By using digital collaboration tools, continuous chats were kept with the stakeholders both individually and in group conversations with people working for the same initiative. The weekly team meetings internally held by the Analytics team moved to online meetings with the same schedule, where the activities were maintained using virtual Kanban boards. There were no further impacts in managing these activities since all the team members worked with these boards before the pandemic.

However, during the pandemic new junior scientists joined the team. Previously, the team integration was held in face-to-face discussions without following a structured agenda -- the required company accesses, team physical layouts and the location of the key contacts in the office usually happened according to the daily schedule of the team members in charge of closely working with new members. The same happened with any technical training required since different data engineer fronts or the CoE had separate software needs. By being completely remote, the new scientists experienced challenges by trying to understand the work processes without comprehensive documentation. In order to mitigate this and as part of another data science project the team created a series of flowcharts to guide newcomers on what are the expected actions to be taken and what kind of machine learning deployment configurations they are expected to do. Also, a formal onboarding presentation had been created containing frequent asked questions, team acronyms, the IT architecture adopted by the team, the different team roles and key contacts as well as the technical skills in use by the team. While some organizations often have extensive documentation by default, it is important to recall the people-oriented instead of process-oriented regional work culture.

\subsection{Practices}
In parallel, the collaboration tools were again used to create team groups with separate channels sharing all the aforementioned documents, best practices, study resources and additional technical resources to connect people. Meetings and team chats do not follow fixed schedules since questions commonly happen in different hours and in separate days. Junior analysts often avoid calls -- especially video calls -- often preferring continuous text communication instead.

A community of practice (CoP) had also been established by the CoE. It is a group of people composed of enthusiasts and stakeholders interested in a specific subject \cite{eckert2006}. As far as the team was aware, there were no active CoPs in the company. This CoP had been named as the \textit{Data Science Network}, and attempted to host webinars on different data science subjects facilitated by members of the CoE AA and presented by either the same members or the data engineers in Analytics. The idea was to engage people from separate areas and expand the CoE influence as positioning itself as the company experts in data science. The Network is composed of members from other IT teams, Analytics and Business teams in different hierarchy levels. The Network includes discussions on industrial data science solutions, tool demos, XAI techniques, programming best practices and performance comparisons. The discussions always attempt to include use-cases from the company itself and comparing against challenges that could happen in the company. It also includes study resources, news from the scientific community and links to upcoming external events such as conferences, for instance.

Although the CoP proved itself to be successful, hosting monthly webinars and in different times to enable the participation of separate people with conflicting schedules, the CoE noticed a need to reach a broader public. Higher management often do not have enough time to participate in the webinars and noted, in separate opportunities, that they prefer content recorded in podcasts. In parallel, Business specialists exposed a need for extensive, continuous learning \cite{peters2017}. Therefore, the CoE also started offering content intended for the high management introducing advanced analytics concepts to them in an industrial context recorded in a podcast. On the other hand, it also was -- on the time this manuscript had been written -- preparing an online course in Python tailored for the Business teams’ requirements. These two examples show positioning opportunities for the CoE: for the former, as a reliable technical advisor team capable of translating highly technical concepts to the decision makers and, for the latter, as a team holding technical expertise and able to successfully share knowledge with other teams. For both cases, the scenario of having data scientists who also worked or were working as professors in universities proved to be a strategic advantage for translating such concepts with empathy, domain of the subject matter and effective communication.

\subsection{Lessons learned}
Over time, the CoE learned the importance of having documented processes to keep basic standards when developing new activities or introducing new members to the team. Also, considering the number of ongoing activities and the lack of face-to-face contact, the team noted it is important to finish the first cycle of the CRISP-DM process as soon as possible. While it might generate a trained machine learning model far from being useful, it helps to attain trust from the stakeholders by proving that a data science project for their use case is indeed possible. However, it is important to ensure the cycle will earn a sense of urgency and prioritization from the stakeholders to have more support from them instead of stalling future cycles by rendering the users comfortable with the first cycle. Therefore, a need of first listing the possible gains generated by the project is relevant to prioritize it and, as a more aggressive measure, to pressure the Business teams to resume the initiative should additional help from these teams is required for steps covered by the methodology such as Data Understanding.

The team also understood the need of having a flexible schedule while also avoiding extremes: a data scientist might spend two full weeks working with data but may spend the following three weeks waiting for the availability of additional data from the data engineers or, still, for answers from the Business teams. In this example, fixed sprint windows as adopted by Scrum teams are not desired.

Understanding the flexibility of work from the data scientists also help them reorganize their upcoming work and working in parallel for several data science projects. The parallelism is helpful considering the Business teams often are unable to provide knowledgeable users full-time, and the number of incoming projects generates the same situation for the data scientists. While quick response times are desired, there might be situations where several projects require the attention of the same data scientists at the same time -- for these cases, prioritizing the projects and discussing with the users in a way they do not feel unattended is important. On the other extreme, large delays in the response from the users is not desired since it could reduce the interest from the Business teams for the project and result in a data science deliverable that is not used by the teams.

The process of building trust with the Business teams is also essential -- in separate feedback sessions some stakeholders highlighted the advantage of working with a data scientist who were both knowledgeable on the Business area and needs, the data science process and the scientific method. Working in the same team of the data engineers also proved to be hugely important due to their shared access and easy reach for discussing and solving issues internally.

\section{Conclusion}
This manuscript proposed to illustrate the context of the data science team of a large company formed by professionals with experience both in industry and academia, demonstrating its challenges and lessons learned during the pandemic.

The team grew organically in over twenty years from being a team of data engineers to an advanced analytics team formed by seasoned data engineers working in Business fronts and data scientists working as a skunk works group. Shortly before the pandemic, the team was starting to have an increased exposure within the organization and implementing new data science projects while it attempted to manage the expectations from the stakeholders as well as pressure coming both internally and externally. Due to its work culture, it embraced the adoption of face-to-face communication instead of an extensive documentation repository. With the pandemic and by forcibly working from home, this model proved as a challenge.

The data science team relied to the adoption of formalized processes and the heavy usage of collaboration tools to keep up with the demand. These tools included messaging and voice applications and digital Kanban boards. The team also noted the advantages of working in parallel in separate projects and noted that junior scientists often favored text communication in the messaging applications instead of voice calls and scheduled meetings. On the other hand, the adoption of a community of practice (CoP), the creation of online courses tailored to the Business teams in the company e-learning platform and podcasts to the higher management helped to foster trust from the stakeholders and ensure they can understand the flexibility of the data science project lifecycle while highlighting the transparency of the process and the shared ownership of the solution between Business and IT, thus nurturing interdependence.

\section*{Acknowledgments}
This study was financed in part by the Conselho Nacional de Desenvolvimento Cient{\'\i}fico e Tecnol{\'o}gico (CNPq), and the Funda{\c c}{\~a}o Arauc{\'a}ria (FAPPR) - Brazil - Finance Codes: 310079/2019-5-PQ2, 437105/2018-0-Univ, 51432/2018-PPP and PRONEX-042/2018.

\bibliographystyle{IEEEtran}
\bibliography{icse_preprint_monteiro_prado}

\end{document}